\documentclass[aps,prb,preprint]{revtex4-1}
\usepackage{amsmath,amssymb,graphicx,bm}
\input epsf

\usepackage{amsbsy}
\usepackage{latexsym}
\usepackage{ulem}
\usepackage{soul}
\usepackage{color}
\usepackage{graphicx}
\usepackage{psfrag}
\usepackage{xfrac}
\usepackage{xfrac}

\begin{document}

\title{An empirical method to characterize displacement distribution functions for anomalous and transient diffusion}
\author{Le Qiao, Nicholas Ilow, Maxime Ignacio, Gary W. Slater}
\affiliation{Department of Physics, University of Ottawa, Ottawa, Ontario K1N 6N5, Canada.}
\date{\today}

\begin{abstract}

We propose a practical empirical fitting function to characterize the non-Gaussian displacement distribution functions (DispD) often observed for heterogeneous diffusion problems. We first test this fitting function with the problem of a colloidal particle diffusing between two walls using Langevin Dynamics (LD) simulations of a raspberry particle coupled to a lattice Boltzmann (LB) fluid. We also test the function with a simple model of anomalous diffusion on a square lattice with obstacles. In both cases, the fitting parameters provide more physical information than just the Kurtosis (which is often the method used to quantify the degree of anomaly of the dynamics), including a length scale that marks where the tails of the DispD begin. In all cases, the fitting parameters smoothly converge to Gaussian values as the systems become less anomalous.
\end{abstract}

\maketitle
\section{Introduction }
\label{section:intro}

Diffusion in inhomogeneous media is ubiquitous in nature and can be observed in a wide range of systems including surface diffusion of atoms in an elastic field \cite{sabiryanovSurfaceDiffusionGrowth2003,xuGrowthKineticsStrained2012,ignacioWettingElasticSolids2014};
diffusion of colloidal particles in a confined fluid\cite{mejia-monasterioTracerDiffusionCrowded2020,sharmaHighprecisionStudyHindered2010,banerjeeExperimentalVerificationNearwall2005,dufresneHydrodynamicCouplingTwo2000} or polymer network\cite{xueDiffusionNanoparticlesActivated2020,singhNonGaussianSubdiffusionSinglemolecule2020};
polymer translocation through a nanopore membrane \cite{dubbeldamDrivenPolymerTranslocation2007,metzlerWhenTranslocationDynamics2003,kantorAnomalousDiffusionAbsorbing2007}; and protein diffusion in crowded cell environments \cite{guigasSamplingCellAnomalous2008,sanabriaMultipleDiffusionMechanisms2007,banksAnomalousDiffusionProteins2005,jeonProteinCrowdingLipid2016a}.
Experimental observations have shown that some of these systems can lead to non-trivial statistical and dynamical properties which require special attention. 

For instance, some systems give rise to anomalous subdiffusion where the mean-square displacement (MSD) increases like $t^\gamma$, with $\gamma <1$. This phenomenon is often related to diffusion in disordered systems, in which case there is a crossover distance beyond which diffusion becomes Brownian ($\gamma=1$). Another particularly interesting case is the possible existence of \textit{anomalous yet Brownian} diffusion which is characterized by a linear MSD ($\gamma=1$) coexisting with a non-Gaussian Displacement Distribution (DispD)\cite{miottoLengthScalesBrownian2021,zhouDiffusionAnisotropicColloids2020,wangUnexpectedCrossoversCorrelated2020,wangAnomalousBrownian2009,wangWhenBrownianDiffusion2012,chechkinBrownianNonGaussianDiffusion2017}. The physical origin of anomalous yet Brownian diffusion remains a very active field of research, particular because it can differ between systems\cite{metzlerSuperstatisticsNonGaussianDiffusion2020}. In the case of heterogeneous diffusion\cite{chubynskyDiffusingDiffusivityModel2014,cherstvyAnomalousDiffusionErgodicity2013a}, the diffusion coefficient is position-dependent and hence varies during the diffusion of a given particle. In the case of models such as the continuous-time random walk\cite{zumofenScaleinvariantMotionIntermittent1993}, on the other hand, diffusion is controlled by a heavy-tailed distribution of waiting times. In both cases, the nature of the DispD is central to our understanding of the physics.

The Kurtosis $K(t)={\mu_4(t)}/{\mu_2(t)^2}$ of the DispD is often used to characterize the deviations from Gaussian (normal diffusion) dynamics\cite{wangUnexpectedCrossoversCorrelated2020,nagaiPositionDependentDiffusionConstant2020,singhNonGaussianSubdiffusionSinglemolecule2020,mejia-monasterioTracerDiffusionCrowded2020} ($\mu_i$ is the $i^{th}$ central moment of the distribution). However, the DispD obviously contains more information than what the Kurtosis provides, including the shape of the tails and potentially the length scale(s) that separate the various dynamical regimes.

In this short article, we propose a new empirical function that can be used to fit a DispD that has different regimes for short and large distances; in particular, it can capture both the Gaussian and non-Gaussian (e.g., exponential) components of a DispD thus allowing us to locate the transition between the two as the external control parameters are changed. To test this flexible interpolating function, we simulate two different systems corresponding to the two classes of problems mentioned above. First, we use a coupled Langevin Dynamics--Lattice Boltzmann (LD-LB) method to simulate the diffusion of a particle in a liquid between two flat walls, an example of anomalous yet Brownian motion\cite{wangAnomalousBrownian2009,hapcaAnomalousDiffusionHeterogeneous2009}. We then study diffusion on a lattice with obstructed sites, a case where short-time diffusion is known to be anomalous \cite{saxtonAnomalousDiffusionDue1994,metzlerRandomWalkGuide2000,jeonProteinCrowdingLipid2016a}.

\section{A practical fitting function}
\label{section:Theory}

To fit both the central and tail parts of "anomalous" DispD distributions at time $t$, we propose to use the 3-parameter functional form
\begin{equation}
    P(r,t) \sim \exp{\left[1-\left[1+\left(\frac{r}{r_o}\right)^{2-\alpha}\right]^{1-\beta}\right]}~,
    \label{eq:generalfct}
\end{equation}
where $r$ is the displacement. The fitting parameters are the length scale $r_o(t)$ and the two exponents, which are expected to be in the ranges $0 \leq \alpha(t) <2$ and $\beta(t)< 1$. A single-regime DispD corresponds to $\beta=0$, including the Gaussian distribution when $\alpha=0$, the exponential function($\alpha=1$), and a stretched exponential distribution ($1 < \alpha <2$). 

This function includes two regimes (which are in fact identical when $\beta = 0$), namely
\begin{equation}
\label{eq:limits}
P(r,t)\sim
\begin{cases}
\exp\left[-(1-\beta)\left(r/r_o\right)^{2-\alpha}\right]  &\text{for } r\ll r_o  \\
\exp\,\left[-\left(r/r_o\right)^{(2-\alpha)(1-\beta)}\right] &\text{for } r\gg r_o.
\end{cases}
\end{equation}The combination $[\beta =\frac{1}{2}; \alpha = 0]$ is a Gaussian with exponential tails, while $1-\beta=\frac{2}{2-\alpha}$ is a distribution with Gaussian tails (a special case being $[\beta =-1; \alpha = 1]$, an exponential distribution with Gaussian tails). Note that $r_o$ marks not only the transition between these two limits, but also the length scale characterizing the decay of the tails of the DispD.


\section{Example I: Wall-Hindered Diffusion}

\label{section:WHD_theory}
Our first example is a simulation of the wall-hindered diffusion of a spherical particle of radius $R$ between two walls separated by a distance $h$. In bulk solution, the diffusion coefficient of the particle is given by Stokes' law $D_0 = {k_BT}/{6\pi \eta R}$, with $\eta$ the viscosity of the fluid. Hydrodynamic interactions (HI) make the diffusion coefficient space-dependent and anisotropic near surfaces. For a particle at a distance $z$ (see Fig.~\ref{Fig:system}a) from a single flat wall \cite{hfaxenFredholmIntegralEquations1924}, the diffusivities parallel and perpendicular to the wall are (with $\Gamma = \frac{R}{R+z}$)
\begin{equation}
\label{eq_D1para}
	 {D_{\parallel}(z)}/{D_0} \approx 1-\tfrac{9}{16} \Gamma + \tfrac{1}{8} \Gamma^3 - \tfrac{45}{256} \Gamma^4 - \tfrac{1}{16} \Gamma^5 + ...
\end{equation}
\begin{equation}
\label{eq_D1perp}
	 {D_{\perp}(z)}/{D_0} \approx (6-10\Gamma+4\Gamma^2)/(6-3\Gamma-\Gamma^2),
\end{equation}

As recently shown by Matse \textit{et al.} \cite{matseTestDiffusingdiffusivityMechanism2017}, the z-dependence of $D_{\perp}$ leads to anomalous yet Brownian motion (linear time-dependence of the MSD but non-Gaussian DispD). 

Since this problem is one-dimensional (along $z$) while $\alpha=0$ (the central part of the distribution is Gaussian\cite{matseTestDiffusingdiffusivityMechanism2017}), the fitting function reads
\begin{equation}
    P(z,t)= \frac{1-A}{\sqrt{2\pi z_o^2}}~ \exp{\left[1-\left[1+\frac{z^2}{2z_o^2}\right]^{1-\beta}\right]}~,
    \label{eq:fct1D}
\end{equation}
where $A$ is a normalization factor. Note that for a Gaussian distribution we simply have $\beta=A=0$, which leads to $\langle z^{2}(t)\rangle=\int_{-\infty}^{+\infty}z^2 P(z,t) \, dz = z_o(t)^{2}=2Dt$, i.e. Brownian diffusion. The even moments of the distribution are given by
\begin{equation}
    \langle z^{2i}\rangle=z_o^{2i} \times \frac{Q(2i,\beta)}{Q(0, \beta)}
\end{equation}
where $Q(2i,\beta)=\int_{0}^{\infty} y^{2i} \exp{\left[1-\left[1+{y}^{2}/2\right]^{1-\beta}\right]} \mathrm{d}y$ and $i$ is an integer. 
In general, both the length scale $z_o$ and the exponent $\beta$ can be time-dependent. The corresponding Kurtosis is given by
\begin{equation}
K(t,\beta)=\frac{\langle z^4(t,\beta)\rangle^{~}}{\langle z^2(t,\beta)\rangle^2}=\frac{Q(4,\beta)~Q(0,\beta)}{Q^2(2,\beta)}~.
\end{equation}
Although $Q(2i,\beta)$ has no closed form, the following second-order approximations are useful for a nearly Gaussian DispD (\textit{i.e.}, when $|\beta| \ll 1$):
\begin{equation}
\label{eq:xo_expand}
   \frac{\langle z^2\rangle}{z_o^2} \approx 1+1.815\beta + 3.121\beta^2+...  
\end{equation}
\begin{equation}
\label{Eq:kurtosis}
  ~~~K-3 \approx 1.0315 \beta +1.588 \beta^2+... ~. ~~~~~~~
\end{equation}
Interestingly, the excess Kurtosis $K-3$ is (to first order) approximately equal to the exponent $\beta$ that characterizes the tail of the DispD. Note that since $\beta$ is expected to be a function of time while $\langle z^2\rangle \sim t$ for this problem, the critical length scale $z_o$ cannot increase like $\sim \sqrt{t}$. Finally, we note that for a Gaussian distribution with perfect exponential tails, our interpolating function predicts that $K(\beta=\frac{1}{2}) \approx 4.857$, which appears to be in agreement with the limit value reported in Ref.\cite{matseTestDiffusingdiffusivityMechanism2017}.

\subsection{Raspberry colloidal particle diffusion}

\begin{figure}[htb]
\begin{center}
\includegraphics[scale=1]{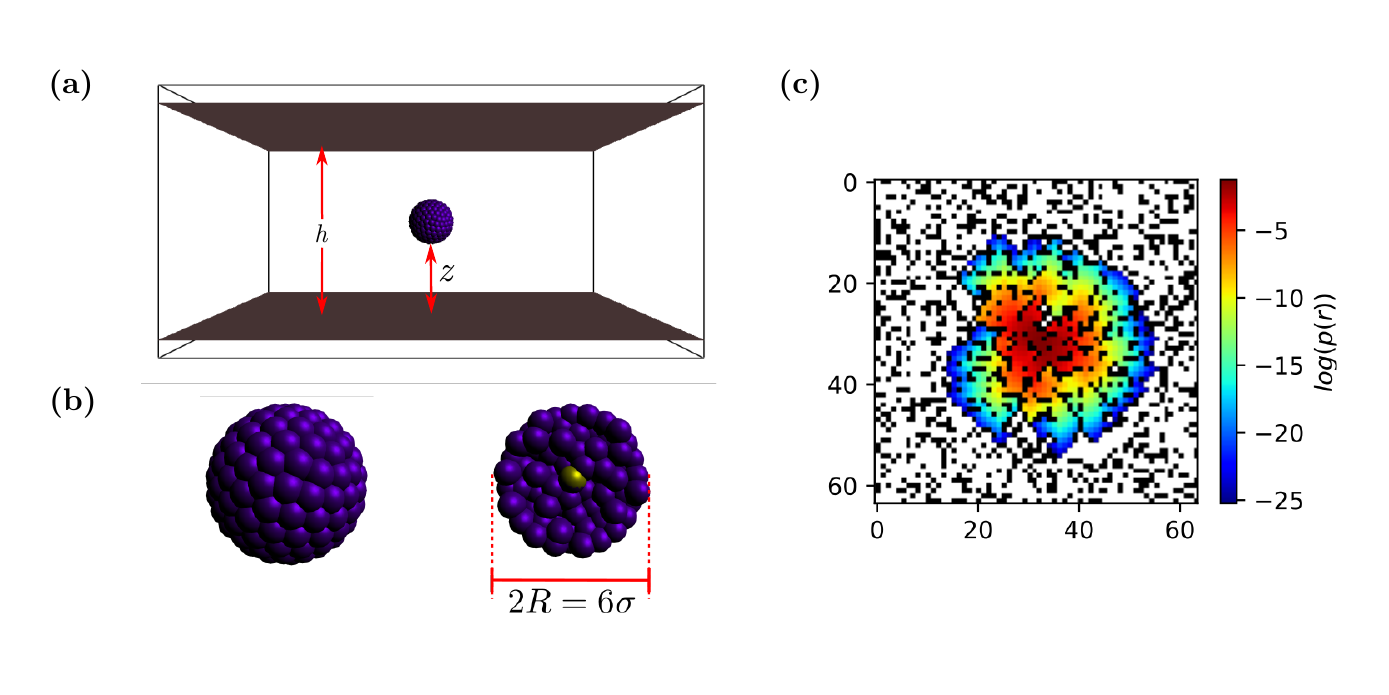}
\end{center}
\caption{The systems used to test the empirical fitting function. (a) Two repulsive boundaries are separated by a distance $h$, while $z$ is the distance between the raspberry particle surface and the wall. (b) A raspberry particle of radius $R=3\,\sigma$ comprising 454 coupled beads. The golden bead is at the center of mass. (c) A two-dimensional 64$\times$64 square lattice with $\phi = 30\%$ of the lattice sites occupied by obstacles (in black). The concentration profile, starting from the site at the center of the system, is shown after 30 time steps.}
\label{Fig:system}
\end{figure}

We use Langevin Dynamics as implemented in the ESPResSo package\cite{weikESPResSoExtensibleSoftware2019}. The HI are included by coupling the particle's velocity to a lattice Boltzmann fluid. We employ the raspberry particle model\cite{fischerRaspberryModelHydrodynamic2015,degraafRaspberryModelHydrodynamic2015,kreisslFrequencydependentMagneticSusceptibility2021} shown in Fig.~\ref{Fig:system}b. The particle comprises $N=454$ beads of size $\sigma$, which gives it a radius $R=3\,\sigma$ and a volume $V_o=\frac{4}{3}\pi R^3\approx113\,\sigma^3$. We freeze all beads relative to the center of mass using virtual momentum-transferring rigid bonds. The particle and the solvent share the same density, $\rho_s=m_o/\sigma^3$, where $m_o$ is the mass of a bead. 

The implicit solvent is modeled by a GPU implementation of the three-dimensional 19 velocity LB method (D3Q19)\cite{dunwegStatisticalMechanicsFluctuating2007}. The velocity $\bm{v}$ of a bead is coupled to the fluid via a drag force\cite{dunwegStatisticalMechanicsFluctuating2007} $\bm{f}_{\gamma,i}=-\gamma (\bm{v}_i-\bm{u}_i)$,
where $\bm{u}_i$ is the velocity of the fluid at the position of the bead (with $i=x,y,z$); the coupling friction $\gamma$ has to be to tuned to insure that the particle's hydrodynamic radius is given by $R_H=R$. 
\begin{sloppypar}
The repulsive Weeks-Chandler-Andersen (W) potential \cite{weeksRoleRepulsiveForces1971} $U_\mathrm{W} (r) =4 \epsilon \left[ \left( \frac{\sigma}{r}\right)^{12} - \left( \frac{\sigma}{r} \right)^6 \right] +\epsilon$ models the steric interactions between a  raspberry bead and a wall when $r<r_c=2^{1/6}\,\sigma$, while $U_{W}(r>r_c)=0$. We use $\epsilon$ for our unit of energy, $\sigma$ for length and $t_o=\sigma\sqrt{m_o/\epsilon\,}$ for time. The temperature is chosen to be $k_BT=\epsilon$ while the kinematic viscosity of the fluid is $\eta=12\,\sigma^2/t_o$ and the coupling per bead is $\gamma=15\,m_o/t_o$. These choices give a hydrodynamic radius $R_H\approx 2.98\,\sigma$ and a diffusion coefficient $D_o=0.00148\,\sigma^2/t_o$ in free solution. We use an integration time step $\delta t=0.005\,t_o$. 
\end{sloppypar}

\subsection{Displacements distributions}

The transverse displacement distribution function $P(z,t)$ (denoted DispD$_\perp$ in the following) is computed for different time intervals and locations between two non-slip walls (Fig.~\ref{Fig:system}b) separated by a distance $h=30\,\sigma$. We start the particle in the center ($z=13.5\,\sigma$) and let it diffuse freely until it reaches $z=\sigma/2$. The simulation times are long enough to generate $2000$ uncorrelated sub-trajectories; we can then use any position in these trajectories as an effective initial position. 

\begin{figure}[thbp]
\begin{center}
\includegraphics[scale=1.2]{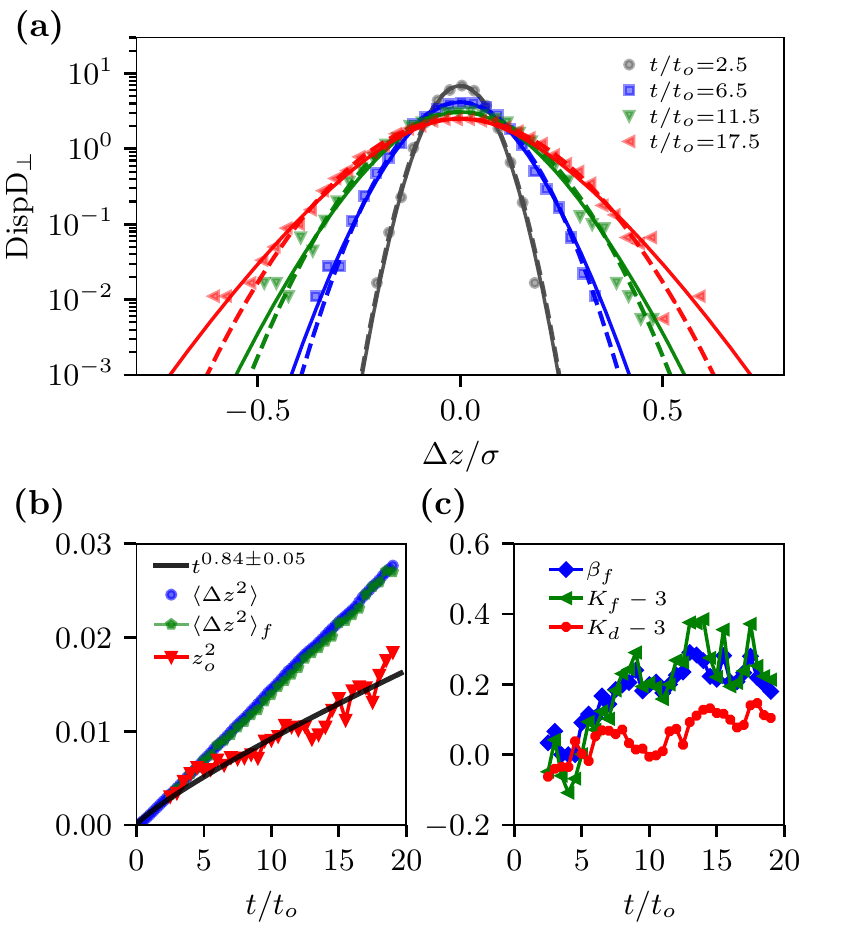}
\end{center}
\caption{Simulation data and fitting parameters for a spherical particle diffusing between two walls. (a) The DispD$_\perp$ averaged over different starting positions $z/R\in[0.5,2]$ at four different times. The simulation data are binned with a bin size $0.03\,\sigma$. The dashed lines are Gaussian fits while the solid lines show the interpolating function given by Eq.~\ref{eq:fct1D}. (b) Length scales as a function of time: $\langle \Delta z^2\rangle$ is the perpendicular MSD calculated from the raw data, while the other two are obtained from the fits. The black curve shows the power law fit $z_o^2 \sim t^{0.84}$. (c) The excess Kurtosis $K_f-3$ and the exponent $\beta_f$ (as obtained from the fits) as a function of time; note that $K_{d}-3$ is calculated directly from the raw data.}
\label{Fig:DispD}
\end{figure}

When a particle diffuses away from a given position for a brief period of time, its diffusivity is essentially constant during the resulting trajectory: our results (Appendix \ref{app-a}) then show that it undergoes normal diffusion (MSD~$\sim t$) but with a local anisotropic diffusivity $D(z)$, in agreement with Eqs.~\ref{eq_D1para}-\ref{eq_D1perp}. However, when the DispD$_\perp$ is averaged over initial positions located in a high $\nabla D_\perp (z)$ region (the diffusing diffusivity regime\cite{chubynskyDiffusingDiffusivityModel2014}), it includes both Gaussian and non-Gaussian components. Figure~\ref{Fig:DispD}a shows examples for starting points located in the region $\tfrac{z}{R}\in [0.5,2]$: the dashed Gaussian fits demonstrate the presence of fat tails, in agreement with ref.\cite{matseTestDiffusingdiffusivityMechanism2017}, while the solid lines show that Eq.~\ref{eq:fct1D} provides excellent fits for all values of the displacement $\Delta z$ and all times. Nevertheless, the MSD is still increasing linearly with time -- see Fig.~\ref{Fig:DispD}b. The DispD remains Gaussian (within the limits of precision we can achieve) in the parallel direction (not shown) because this diffusivity gradient is weaker.

As previously mentioned, a common way to characterize an "anomalous DispD" is to compute its Kurtosis. Figure~\ref{Fig:DispD}c shows that the excess Kurtosis (both of the fitted function, $K_f-3$, and of the raw data, $K_d-3$) vanishes at short times (we then probe local regions with uniform diffusivity). We note that $K_f>K_d$ because the tails of the raw data distributions are heavily truncated here (at distances comparable to the length scale $z_o$). 

Our fitting function includes additional information. For instance, we see that the exponent $\beta$ mirrors the behavior of the Kurtosis. In fact, we note that $K_f(t)-3 \approx \beta(t)$ when $\beta$ is small in Fig.~\ref{Fig:DispD}c, in agreement with Eq.~\ref{Eq:kurtosis}. The fit also provides information about the length scale $z_o$ beyond which the Gaussian part (short distances) changes to fat tails (large distances): as shown in Fig.~\ref{Fig:DispD}b, our results indicate that $z_o^2$ increases roughly as $\sim t^{0.84}$ over the time range used here. The fitting parameters thus show that the reason why the MSD increases linearly with time is that the sublinear increase of $z_o^2$ is compensated by an increase of the anomalous exponent $\beta$ over the time periods studied here.

Recently, a similar problem involving non-spherical molecules instead of particles was studied in ref \cite{fernandezDiffusionDoxorubicinDrug2020} using all-atoms molecular dynamics computer simulations. It would be interesting to see if our fitting function could be used instead of their Eq.\,6 since the exponent in their stretched exponential was different for short and large distances, something that our function can easily accommodate.

\section{Example II: Anomalous diffusion in random systems}

We now examine the usefulness of the general form given by Eq.~\ref{eq:generalfct} for a case of obstructed diffusion. We use the simplest model: a random walk on a two-dimensional square lattice with a fraction $\phi$ of the sites being randomly occupied by immobile obstacles (see Fig.~\ref{Fig:system}c). In short, diffusion is expected to be normal (\textit{i.e.}, the MSD grows linearly with time and the DispD is Gaussian) for short times (before the particle starts colliding with the obstacles; in practice, this regime only exists at very low obstacle concentration) and long times (the steady-state, achieved for times larger than the crossover time $t^*$ and distances larger than the system's crossover length $r^*$). For intermediate times, the MSD grows roughly as $t^\gamma$, where the anomalous subdiffusion exponent $\gamma < 1$. See refs. \cite{saxtonAnomalousDiffusionDue1994,slaterExactlySolvableOgston1996} for previous studies of this system. Obviously, given the transient (or anomalous) regime mentioned above, we expect non-Gaussian radial distribution functions $P(r)$ (where $r^2=x^2+y^2$) unless $t \gg t^*$.

\subsection{Random-Walk on a 2D Lattice}

We randomly place a concentration $\phi$ of $1 \times 1$ obstacles on a $1600 \times 1600$ square lattice, and average over an ensemble of $500$ different obstacle configurations for each value of $\phi$. In order to obtain the radial distribution function $P(r,t)$ at different times $t$ we use a standard Markov Chain Monte Carlo propagation algorithm: we initially place a unit concentration on the lattice site at the center of the system, and we propagate the concentration throughout the lattice using the master equation. The jumping probabilities are chosen to be $p_{\pm x}=p_{\pm y}=p=\frac{1}{8}$; the  probability of not jumping at each time step is thus $p_0=\frac{1}{2}$. The free diffusion coefficient is set to $D=1$, and the lattice spacing is set to $a=1$. The time duration of a Markov step is then simply $\tau = \frac{a^2p}{D} = \frac{1}{8}$. The resulting radial distribution functions (a few examples are shown in Fig.~\ref{Fig:NickDists}), at specific times, are then fitted with both a Gaussian and the fitting function 
\begin{equation}
\label{eq:2dfit}
    P(r)= \frac{2(1-A)}{\sqrt{\pi r_o^2}}~ \exp{\left[1-\left[1+\left(\tfrac{r}{r_o}\right)^{2-\alpha}\right]^{1-\beta}\right]}~
\end{equation}
(the Gaussian distribution corresponds to $A=\alpha=\beta=0$). Once the fitting parameters are obtained, it is possible to compute the even moments of the radial displacement
\begin{equation}
    \langle r^{2i}\rangle=r_o^{2i} \times \frac{G(2i,\alpha,\beta)}{G(0,\alpha, \beta)}
\end{equation}
where $G(2i,\alpha,\beta)=\int_{0}^{\infty} y^{2i+1} \exp{\left[1-\left[1+{y}^{2-\alpha}\right]^{1-\beta}\right]} \mathrm{d}y$. 
In principle, the only explicit time-dependence is in the length scale $r_o(t)$, with $r_o^2(t)=\langle r^2(t)\rangle=4Dt$ for normal diffusion in two dimensions; however, the exponents $\alpha$ and $\beta$ can also vary with time, as we shall see. The 2D Kurtosis is then given by
\begin{equation}
    K_2(\alpha,\beta)=\frac{G(4,\alpha,\beta)~G(0,\alpha,\beta)}{G^2(2,\alpha,\beta)}~.
\end{equation}
Once more, although $G(2i,\alpha,\beta)$ has no closed form, the following approximations are useful for a nearly Gaussian DispD (\textit{i.e.}, when $\alpha \ll 1$ and $|\beta| \ll 1$):
\begin{equation}
\label{eq:r2_expand}
   \frac{\langle r^2\rangle}{r_o^2} \approx 1+0.7114 \, \alpha + 2 \, \beta+...  
\end{equation}
\begin{equation}
\label{Eq:kurtosis2}
  ~~~~K_2\approx 2+\tfrac{1}{2} \, \alpha +0.5963 \, \beta+... ~. 
\end{equation}
Interestingly, the excess Kurtosis $K_2-2$ is (to first order) approximately equal to the mean value of the two exponents, $(\alpha+\beta)/2$. It is also possible to compute the moments of the distribution itself (treating the problem as a one-dimensional distribution). We then obtain the following 1D Kurtosis:
\begin{equation}
\label{Eq:kurtosis3}
  K_1 = \frac{G(3,\alpha,\beta)~G(-1,\alpha,\beta)}{G^2(1,\alpha,\beta)} \approx 3+ \alpha +1.0315 \, \beta+... ~. 
\end{equation}
Note that the excess Kurtosis $K_1-3$ is (to first order) approximately equal to sum of the two exponents, $\alpha+\beta$ here. We will be using $K_2$ in our analysis below.

\subsection{Displacement distributions \textit{vs} time}

\begin{figure}[thpb]
    \begin{center}
        \includegraphics[scale=1.2]{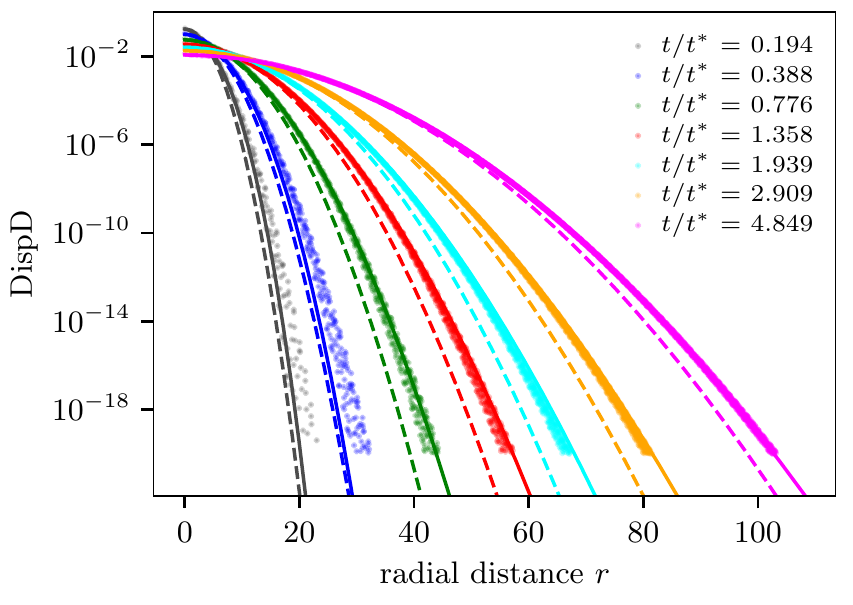}
    \end{center}
\caption{The radial displacement distribution function DispD at six different times for the two-dimensional obstructed diffusion problem. The simulation data are not binned (every possible radial distance $r$ on a square lattice is considered, hence the apparent noise in the otherwise exact data). The dashed lines are fits while the solid lines show the interpolating function given by Eq.~\ref{eq:2dfit}. The obstacle concentration is $\phi = 1/9$. Note that the transient/anomalous exponent is $\gamma=0.963$ for this system, while the crossover length and time are $r^* = 6.95$ and $t^* = 16.11$, respectively.}
\label{Fig:NickDists}
\end{figure}

Our fitting function does not work well at short times in Fig.~\ref{Fig:NickDists} because of the coarse lattice discretization effects for short displacements (although it performs slightly better than the Gaussian function). Once we approach the crossover time $t^*$ marking the transition between the transient and steady-state regimes (see ref. \cite{ilowEstimatingSteadyState2021} for details regarding the method used to obtain $t^*$ from the data), the tails are very well captured by our function while the Gaussian fits are clearly inadequate. As expected, the Gaussian fits converge with the data only when we are well into the steady-state, i.e. when $t \gg t^*$.

\begin{figure}[thpb]
    \begin{center}
        \includegraphics[scale=1.2]{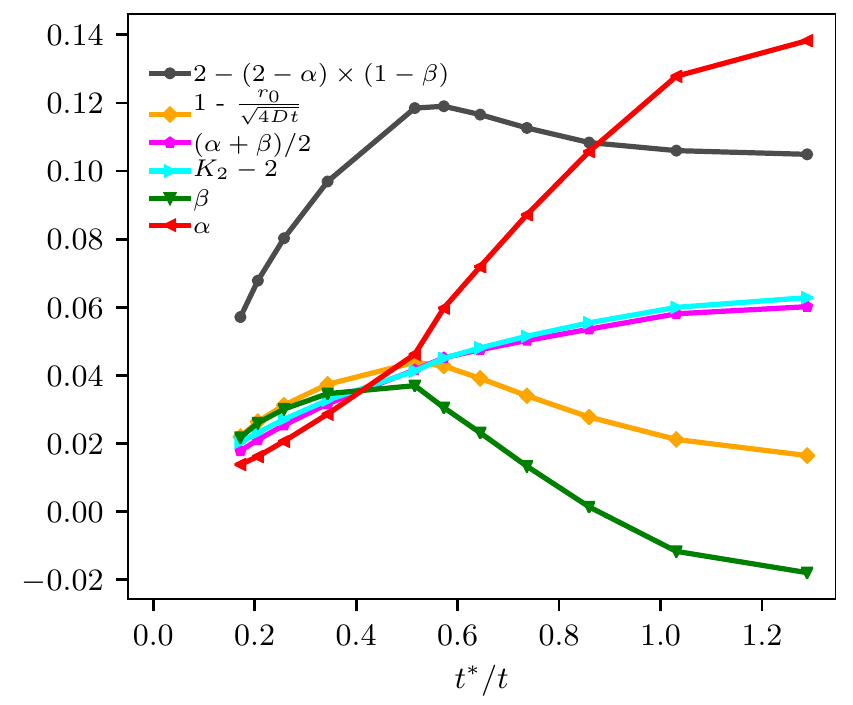}
    \end{center}
\caption{Fitting exponents $\alpha$ and $\beta$, from Eq.~\ref{eq:2dfit}, \textit{vs} inverse rescaled time $t^* / t$. We also show that the Kurtosis of the distribution, $K_2$, and the product $(2-\alpha)(1-\beta)$ both converge to the Gaussian value of 2 at long time. The average of the exponents $\alpha$ and $\beta$ provides a good approximation for $K_2 - 2$. The fitting length scale $r_o(t)$ approaches the diffusion length $\sqrt{4Dt}$ in the asymptotic limit.}
\label{Fig:Nick_Params}
\end{figure}

We examine how the fitting parameters evolve as a function of inverse scaled time $t^*/t$ in Fig. \ref{Fig:Nick_Params}. As discussed above, we expect the distribution to become more Gaussian (which means the exponent characterizing the tail of the distribution, $(1-\beta)(2-\alpha)$, converges towards $2$; see Eq.~\ref{eq:limits})  as time increases, and fully Gaussian distributions (\textit{i.e.}, $\alpha=\beta=0$) for times $t \gg t^*$. Indeed, these parameters, as well as the Kurtosis, converge towards their Gaussian limits when $t \gg t^*$ in Fig.~\ref{Fig:Nick_Params}. This agrees with Fig.~\ref{Fig:NickDists} where it is clear that the fits become more Gaussian as $t/t^*$ increases. We also note that the excess Kurtosis $K_2-2$ is essentially given by the mean $(\alpha+\beta)/2$ over the whole time interval.

However, the distribution function is not expected to be Gaussian over radial distances $r<r^*$ (or times $t < t^*$) since diffusion is anomalous over these length scales (\textit{i.e.}, we then have $\langle r^2(t) \rangle \sim t^\gamma$, with $\gamma=0.961$ here; data not shown). In Fig.~\ref{Fig:Nick_Params} we see two transitions in the time dependence of the fitting exponent $\beta$, namely a change in the sign of the slope at $t^*/t \approx 0.5$, and a change in the sign at $t\approx t^*$. The value of $\alpha$, on the other hand, remains positive and decreases very rapidly with time. We also note that the product $(2-\alpha)(1-\beta)-2 \approx -0.05$ is roughly constant for $t<2t^*$, implying that the tail of the distribution decays roughly as $\sim \exp[-(r/r_o)^{1.95}]$ during the anomalous/transient regime. In fact, the time dependence of all the parameters shown here changes in a qualitative way at $t\approx 2 t^*$, except for the Kurtosis. The fit thus provides additional information about the transition between anomalous and steady-state diffusion.

Of great interest is the length scale $r_o(t)$ that characterizes the tail of the radial distribution (see Eq.~\ref{eq:limits}). In Fig.~\ref{Fig:Nick_Params} we renormalize the value of $r_o(t)$ by the long time Gaussian limit $r_0(t) = \sqrt{4Dt}$. The ratio $r_o(t)/\sqrt{4Dt}$ should converge towards unity when time increases: this is precisely what the diamond (orange) data points show. At short times, however, we expect anomalous diffusion with $\left< r^2(t) \right> \sim t^\gamma$, hence the non-monotonic behavior. Finally we find that the characteristic length scale $r_o(t)$ is equal to the crossover length $r^*$ at $t \approx 1.06\, t^*$, showing the connection between the anomalous diffusion regime and the tails of the DispD.

\section{Conclusion}
\label{section:conclu}

In this article, we propose a new interpolating function that can conveniently characterize displacement distributions that contain both Gaussian and non-Gaussian exponential components. The key advantage of our function is that it can describe distributions that have core and tail components with different behaviours, and yet it includes only two additional fitting parameters (compared to the Gaussian fit). In particular, compared to the often used Kurtosis analysis, our method provides more information, including the length/time scale(s) separating the different regimes.

We tested our interpolating function using two simple examples. In the first test, the physics of the problem is such that the distribution has to be Gaussian at short distances, but may have non-Gaussian tails. In the second test, the radial distribution is more general and the fits provided detailed information about its time evolution. We are currently studying how the fit parameters (especially the length scale $r_o$ and tail exponent $(2-\alpha)(1-\beta)$) are connected to fundamental elements of the physics of these two problems such as the local diffusivity gradient (in the case of the first example) and the anomalous diffusion exponent $\gamma$ (the second example). It is interesting to note that a recent paper by Miotto et al. \cite{miottoLengthScalesBrownian2021} discusses the time-dependence of the length scale $\lambda$ related to the tail of the DispD for a problem of "Brownian yet non-Gaussian dynamics"; their main conclusion is that $\lambda$ does not scale diffusively with time, just like $z_o$ and $r_o$ in our two examples.

Since our fitting function was not designed for any specific type of anomalous diffusion, it should prove a useful tool to characterize non-Gaussian displacement distribution functions in a wide range of diffusion systems beyond the two examples discussed in this short paper, including, for example, the anomalous diffusion of active cells \cite{jeonProteinCrowdingLipid2016a}. The characterization and identification of anomalous diffusion remain a challenging task, as recently described in Ref. \cite{munoz-gilObjectiveComparisonMethods2021b}; this is especially true if we have to distinguish between different types of anomalous diffusion using a limited number of short trajectories. To this end, it would interesting to test our fitting function together with machine-learning-based approaches \cite{kowalekClassificationDiffusionModes2019,argunClassificationInferenceSegmentation2021a,janczuraClassificationParticleTrajectories2020b}.

\begin{acknowledgments}
GWS acknowledges the support of both the University of Ottawa and the Natural Sciences and Engineering Research Council of Canada (NSERC), funding reference number RGPIN/046434-2013. LQ is supported by the Chinese Scholarship Council and the University of Ottawa. NI is partially supported by the University of Ottawa.
\end{acknowledgments}

\appendix

\section{Diffusion coefficient at different heights}
\label{app-a}

\begin{figure}[htpb]
\begin{center}
\includegraphics[scale=1.4]{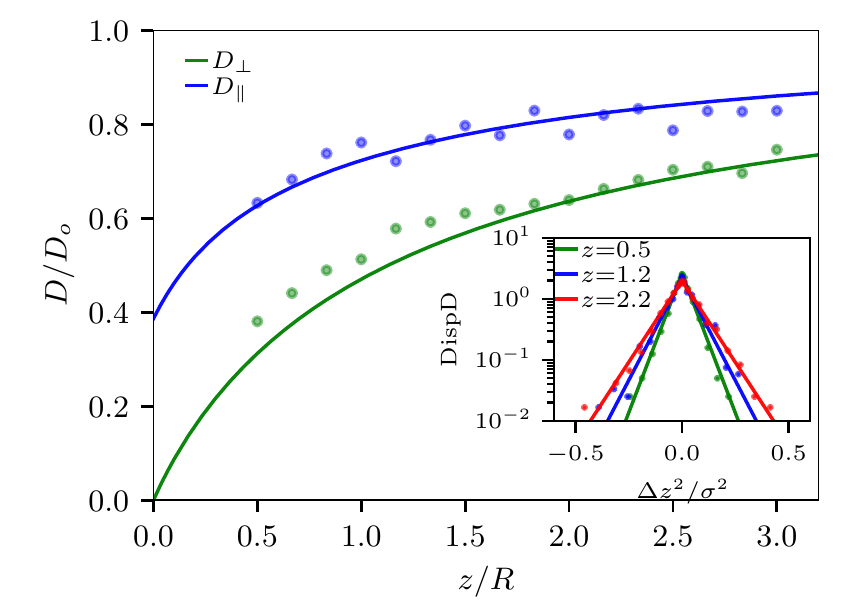}
\end{center}
\caption{Reduced diffusion coefficients parallel ($D_{\parallel}/D_o$) and perpendicular ($D_{\perp}/D_o$) to the wall \textit{vs} the vertical distance $z/R$ from the closest wall, where $D_o=0.00148\,\sigma^2/t_o$ is the free solution diffusion coefficient of the spherical particle. The data points are obtained by fitting the time dependence of the MSD between $t=0$ and $t=20\,t_o$. The solid lines are theoretical predictions given by Eqs.~\ref{eq_D1para} and \ref{eq_D1perp}. Inset: Gaussian fit of the vertical DispD$_\perp$ for three different initial positions $z/R$=0.5, 1.2 and 2.2.}
\label{Fig:Dz}
\end{figure}

Figure~\ref{Fig:Dz} shows the diffusion coefficients $D_\perp$ and $D_\parallel$ measured at different distances $z$ from the wall. Our data are in a good agreement with theory (Eqs.~\ref{eq_D1para} and \ref{eq_D1perp}) for a spherical particle, thus validating our simulation approach.

The inset shows that we indeed obtain a Gaussian DispD$_\perp$ for $t=12.5\,t_o$, with a variance $\sim D_\perp t_o$, for different initial positions $z/R$.

\bibliography{Ref1}

\end{document}